\documentclass[11pt]{article}

\usepackage[style=authoryear]{biblatex}
\usepackage{hyperref}
\hypersetup{colorlinks,allcolors=black}
\usepackage{xcolor}
\usepackage{soul}

\usepackage{graphicx} 
\usepackage[T1]{fontenc}
\usepackage{setspace}
\usepackage[letterpaper, portrait, margin=1in]{geometry} 

\title{\Large{\textbf{Value, Representation, Information and Communication}}}

\author{Xiangjun Peng\\
Succincter\\
\href{mailto:shiangjun@succincter.com}{shiangjun@succincter.com}\\
}

\date{}

\begin{document}

\maketitle

\begin{abstract}
    \noindent
    A mathematical framework, originally intended for information communication, is first described by~(\cite{BellTechn48/Shannon}). The formalization also implies an underlying assumption: all information is disordered. Intuitively, such an assumption shall be considered reasonable. However, a concrete example is demonstrated by~(\cite{ArXiv24/Peng}), which highlights the merits of ordered values for information representations. Therefore, such a contradiction motivates a retrospection on the relationships among value, representation and information; and further investigations, on how their newly-formalized relationships influence communication, are needed.\\

    \noindent
    The basis of this work is a new analytic framework to address the above issue. The framework is first formalized via the usage of the Monadology~(\cite{leibniz/monadology}), to expand the understanding of Zermelo$\hyphen$Fraenkel$\hyphen$choice set theory (ZFC) and Von Neumann$\hyphen$Bernays$\hyphen$G\"odel set theory (NBG). Implicitly, the framework levels value, representation and information separately. Given the fact that there exists a coincidental equivalence between Von Neumann universe and originally-formalized motivation in ZFC, this work hypothesizes the essential of ordered values for one monand, to carry out efficient communication with the rest. \\
    
    \noindent
    This work then focuses on the relationship among values, representation and information (and suggests potential methods for quantitative analysis). First, this framework generalizes the definition of values and representations from ``Indexes $\approx$ Values" principle by~(\cite{ArXiv24/Peng}) via surreal numbers~(\cite{knuth/surreal-numbers}). Second, credited to surreal numbers, this work recursively connects representations and information via subsets of sets. Therefore, the definition to metric space(s) is naturally formed by representations, and quantitative methods (e.g., Hausdorff Distance) can be applied for quantitative analysis among (sub)sets. Third, this framework conjectures that: as long as the metric space is (or can be formed as) complete, the existence tests can be performed via Cauchy Sequence (or its generalized methods).\\

    \noindent
    This work finally revisits the communication theory, and suggests new perspectives from the new analytic framework. First, this work hypothesizes a (quantitative) relationship between values and representation, and conjectures that: the optimal construction of representations exists, and it can be derived as the core value of one monad via Cauchy Inequality (or its generalized methods). Second, this work identifies that: the differences among value sets from different monads are not obstacles for efficient communications, as long as the core values are shared by different monads. Third, efficient communication can be translated into the better efficiency via functionality agreements among expected monads, who share the same core values.\\

\end{abstract}

\newpage
\tableofcontents
\newpage

\begin{singlespace}

\section{Introduction}

\vspace{-4pt}

Information communication is originally formalized by~(\cite{BellTechn48/Shannon}) in a mathematical framework, which introduces a methodology for the quantification. A subsequent line of efforts are made to improve the communication efficiency via such a quantification, and this is due to the limit of lossless compression is expected to exist. However, a recent work by~(\cite{ArXiv24/Peng}) demonstrates that: it is feasible to perform computation directly on the limit of such a compression. Therefore, a retrospection on how we understand information and communication is indeed needed.

In the original formalization from~(\cite{BellTechn48/Shannon}), the main focus lies on solely the transmission. Motivated by~(\cite{FOCS08/Succincter}), the redundancy can improve the functionality efficiency directly on lossless compression. Therefore, this work takes a different approach from~(\cite{BellTechn48/Shannon}): instead of solely focusing on the transmission, this work leverages the concept of Monadology~(\cite{leibniz/monadology}), and assumes all communications are performed among different monads.

This work first employs set theory for information formalization, and introduces monads to differentiate the outstanding differences between Zermelo-Fraenkel-choice set theory (ZFC) and Von Neumann$\hyphen$Bernays$\hyphen$Gödel set theory (NBG). More specifically, this work leverages the viewpoint from Gödel, who assumes the source of NBG comes from a self-centered monad\footnote{Note that the author misses the exact citation, but he recalls that this shall originally come from Gödel's conversation with his PhD advisor.}. Therefore, this work classifies monads into two classes: active monads, which made their own choices; and reactive monads, which obtained labels from others (i.e., namely, class labels within NBG).

Based on the assumption of active and reactive monads, this work leverages the coincidental equivalence between Von Neumann Universe (formulated in NBG) and the originally-formalized motivation from ZFC, and conjectures that: both approaches are feasible to achieve the well-ordered values for one monad. Furthermore, this work makes the hypothesis that: the ideal pre-assumption for efficient communication is to obtain Von Neumann Universe for one monad. Following the formalization of set theory for information, this work takes advantage of the ``Indexes $\approx$ Values" principle~(\cite{ArXiv24/Peng}), and generalizes the results via surreal number~(\cite{knuth/surreal-numbers}). More specifically, within the definition of~(\cite{ArXiv24/Peng}), the values can be represented via surreal number and such results can be recursively connected with~(\cite{ArXiv24/Peng}) and the above formalization. 

Note that the above formalization in terms of value, representation and information brings additional benefits in terms of quantitative analysis. This work highlight some key components, which the author assumes important. First, the metric space(s) can be defined among (sub)sets, and therefore the distances can be defined using Hausdorff Distance. This can contribute to precision control, which can be used to adjust the value-information relationships; and second, the existence test can be carried out via Cauchy Sequences (or its generalized methods), since the results from~(\cite{ArXiv24/Peng}) also suggest the equivalence between information and functionalities.

This work finally revisits the communication theory, and suggests new perspectives from the new analytic framework. First, this work conjectures that: the optimal construction of the representation from one monad exists, and it can be derived via Cauchy Inequality (or its generalized methods). Second, this work identifies that: the differences among value ranges from different monads shall not be obstacles for efficient communications, as long as the core values are shared. Third, it is essential to obtain functionality agreements among monads, for the better efficiency.\\

\vspace{-14pt}

\begin{quote}
    \textit{“The larger the mass of collected things, the less will be their usefulness. Therefore, one should not only strive to assemble new goods from everywhere, but one must endeavor to put in the right order those that one already possesses.”}\\
    \vspace{-16pt}
    \flushright \textit{- Gottfried Leibniz}
\end{quote}

We first give an overview of the results from this work. Then we summarize the techniques developed in this work.

\vspace{-4pt}

\subsection{Our Results}

\vspace{-4pt}
\subsubsection{Formalization on Value, Representation and Information}

\vspace{-4pt}
\noindent
This work first suggests a comprehensive formalization to consistently connect value, representation and information via set theory. Such a consistency via set theory allows quantitative analysis from (ordered) value, representation (if needed), and (disordered) information. Such a formalization also equalizes the Von Neumann Universe as the most optimal recognition from one monad, and this shall correspond to viewpoints on perfection and imperfection from Theodicy~(\cite{leibniz-essais}).

\vspace{-4pt}
\subsubsection{A New Analytic Framework with Quantitative Methods}

\vspace{-4pt}
\noindent
This work then delivers a new analytic framework, which have the following important insights. First, the new analytic framework allows any problem formulations with a given range of values, in a way that value precision and distance metric(s) among (sub)sets can be connected: this allows reactive adjustments for a variety of problem formulations. Second, the new analytic framework bridges (ordered) value, representation (if needed) and (disordered) information: this allows existence tests via Cauchy Sequences (or its generalized methods).

\vspace{-4pt}
\subsubsection{Implications on (Information) Communication}

\vspace{-4pt}
\noindent
This work finally revisits the communication theory, with three new insights from the perspective from monads rather than the transmission. There are two important results. First, the optimal construction of the representations can exist, and would be delivered via Cauchy Inequality (or its generalized methods). Second, the functionality agreements among monads are the preliminary for the efficient communication, as long as the core values are shared.

\vspace{-4pt}
\subsection{Technique Outline}

    \noindent
    There are the following techniques derived from this work, which are considered novel to the literature, and potentially useful for further usage.\\

    \noindent
    \textbf{$\bullet$ Consistently-Recursive Formalization among Value, Representation and Information}: the first novelty of this work is from the generalization of value in ``Indexes $\approx$ Values" via surreal numbers, and connects such a generalization with information via set theory.  This hypothesis allows a consistent usage of set theory throughout value, representation and information.\\

    \noindent
    \textbf{$\bullet$ Value Precision versus Distance Metric(s)}: the second novelty of this work is from the bridge between value precision and distance metric(s). Such an interesting formalization allows dynamic adjustments of value precision and/or distance metric(s), which can be considered reactive (depending on problems). Moreover, these usages can also enable quantitative methods for particular problem settings (e.g., existence tests).
    \\

    \noindent
    \textbf{$\bullet$ New Guidelines on Efficient Communications}: the third novelty of this work is a different viewpoint on communication. Rather than the focus on transmission~(\cite{BellTechn48/Shannon}), this work highlights two parts for efficient communications: (1) the potential for the optimal construction of representations (via Cauchy Sequences) from one monad; and (2) the importance of functionality agreements among different monads, who share the core values. 

\section{Background and Motivation}

The mainstream understanding of information is derived from~(\cite{BellTechn48/Shannon}), which is originally intended for communication.

\vspace{-4pt}

\subsection{Shannon's Mathematical Framework of Communication}

Since the original focus of~\cite{BellTechn48/Shannon} is on communication, data transmission is the major consideration point. Hereby, a figure gives out the concept pictorially.

\vspace{-2pt}

\begin{figure}[!h]
    \centering
    \includegraphics[width=0.7\linewidth]{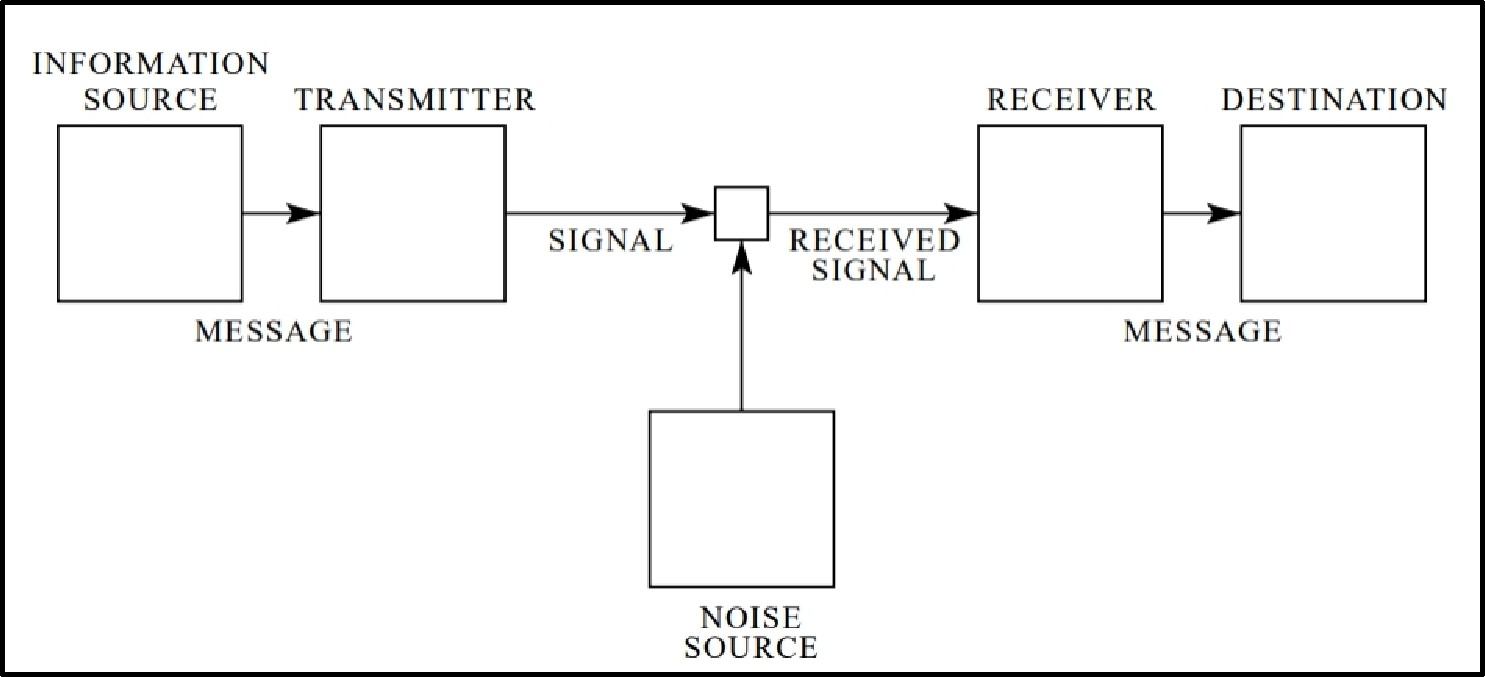}
    \caption{A conceptual formalization of Shannon's mathematical framework for communication from~(\cite{BellTechn48/Shannon}).}
    \vspace{-4pt}
\end{figure}

\vspace{-8pt}

\subsection{Motivation and Methodology}

\vspace{-4pt}

\subsubsection{Outstanding Issues with Shannon's Framework}

\vspace{-4pt}

Data transmission is not equivalent to communication, and this work first document the following outstanding issues from~\cite{BellTechn48/Shannon}, which serves as the motivation of a revisit to information (and communication) in this work.

(1) The focus on information (or representations) overlooks the underlying values. Such a overlook can cause the loss of the efficiency in potential. This can be evidenced by the successful breakthrough from~(\cite{ArXiv24/Peng}), which equalizes compressed indexes and values for efficient queries and computation on near-optimal lossless compression.

(2) The lack of considerations on dimensional aspects for information (e.g., functionalities, time and space). Leveraging the results from~(\cite{ArXiv24/Peng}), the transmitted information can be used for queries and computation, rather than simply being transmitted between sources and destinations ((and neighborhood implications, if there are multiple).

(3) The overlook of the expected functionalities from source/destination. All transmitted information does NOT consider the expected functionalities from source/destination. Therefore, the such a transmission can be useless if there are no clear agreements.
\vspace{-4pt}

\subsubsection{Monadology: A New Perspective}
\vspace{-4pt}

Rather than focus on the transmission, this work leverages Monadology~(\cite{leibniz/monadology}) to revisit the information (and communication) theory. A monad is defined as an elementary individual substance, which reflects the order of the world and from which material properties are derived. Therefore, there are two parts to examine: (1) how can a monad be organized in terms of value, representation and information? and (2) how can monads communicate with each other efficiently?

\section{Value, Representation and Information}

\vspace{-4pt}

\subsection{A Consistently-Recursive Formalization}

\vspace{-4pt}

This work aims to bridge (ordered) value, representation (if needed) and information. This work begins with ``Value versus Information", then discusses how representations play a role, if needed.

\noindent
$\bullet$ \textbf{\underline{Value versus Information:}} this work suggests surreal number~(\cite{knuth/surreal-numbers}) as the generalized definition of value; and then leverage the concepts from set theory for information formalization. Moreover, this work considers one monad is active under Zermelo$\hyphen$Fraenkel$\hyphen$choice set theory (ZFC); and then consider one monad is reactive under Von Neumann$\hyphen$Bernays$\hyphen$Gödel set theory (NBG). Hereby, the formalization is consistently recursive.

\noindent
$\bullet$ \textbf{\underline{Representation:}} this work then considers representation as the bridge between value and information (if needed). The representation can be understood as the alphabet/dictionary, which organizes value in an elementary manner of information. As demonstrated in~(\cite{ArXiv24/Peng}), it is feasible to equalize value (for computation) and information (for queries): in other words, representations can be bypassed/ignored during the execution.

\noindent
$\bullet$ \textbf{\underline{Optimal Cognition from One Monad:}} this work considers the Von Neumann Universe as the optimal cognition from one monad.

\vspace{-4pt}

\begin{figure}[!h]
    \centering
    \includegraphics[width=0.3\linewidth]{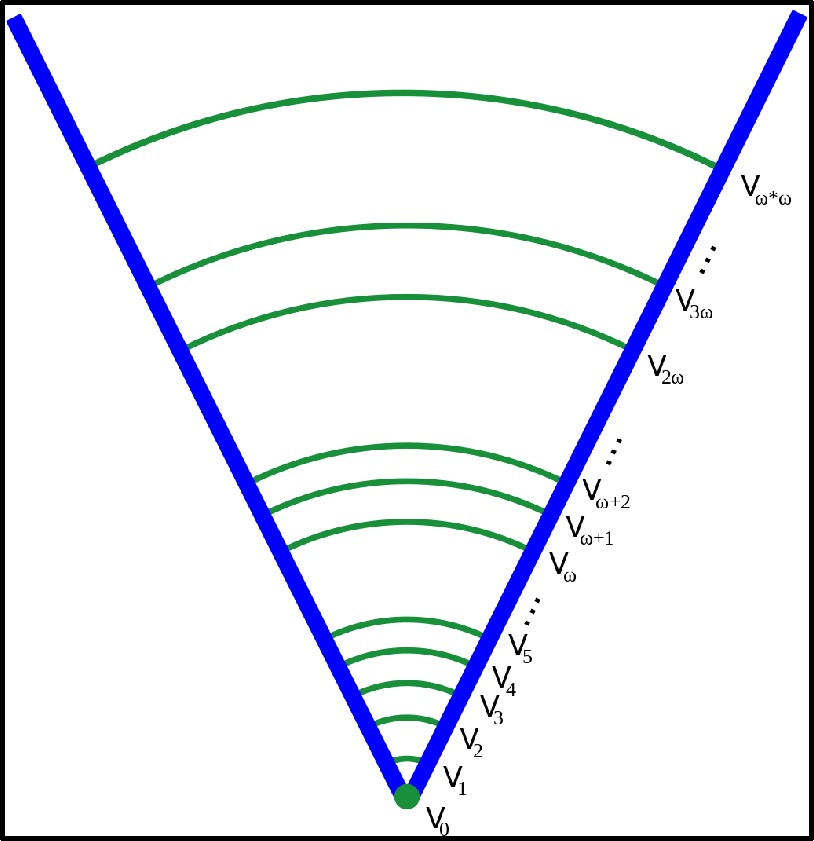}
    \includegraphics[width=0.6\linewidth]{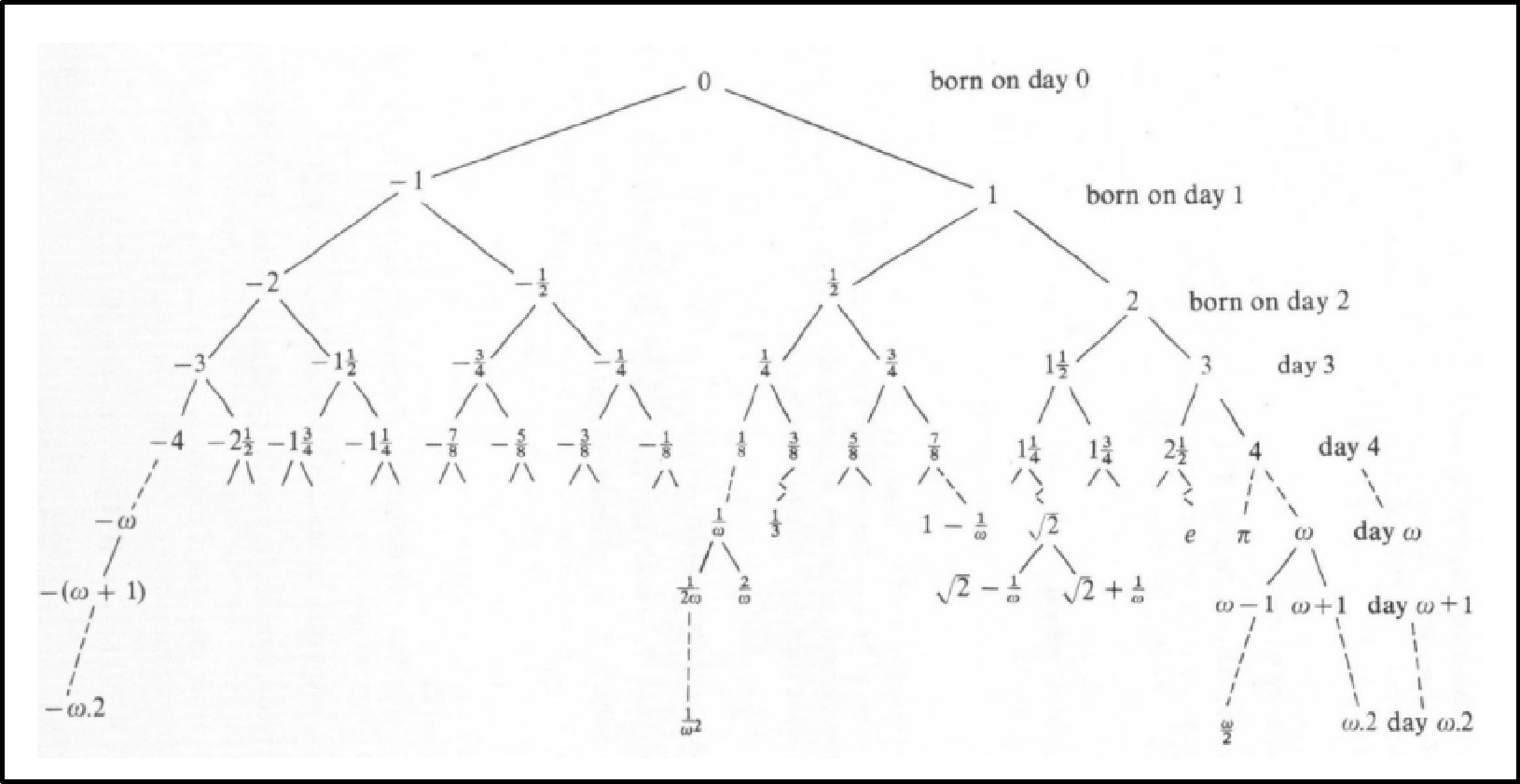}
    \vspace{-8pt}
    \caption{A pictorial example of Von Neumann Universe (left); and a pictorial example of surreal number from Gretchen Grimm on ``Day 0 to infinity" (right).}
\end{figure}

\vspace{-18pt}

\subsection{Takeaways from the Formalization}

\vspace{-4pt}

There are a few notes, which this work assumes important, to be highlighted.

First, given the fact that there exists a coincidental equivalence between Von Neumann Universe and originally-formalized motivation in ZFC, this work conjectures that both active monad (driven by choices) and reactive monad (driven by labels) can reach to the optimal cognition, though the maximum cognition of a monad is always limited.

Second, the described formalization differentiates the perfection and imperfection~(\cite{leibniz-essais}); and the focus of NBG on a self-centered monad also implies the known and the unknown of one's cognition. Second, the formalization via monads also implies the essentials of self-retrospection: namely how to (1) select the needed information for itself; and (2) distill the underlying values from the representations of the received information.

\vspace{-8pt}

\section{A New Analytic Framework with Quantitative Methods}

\vspace{-8pt}

With the above formalization, a new analytic framework is delivered for value, representation and information. This also comes with the quantitative methods, for examining (ordered) value, representation (if needed) and (disordered) information.

\vspace{-8pt}

\subsection{Precision Control, Distance Adjustment and Existence Test} 

\vspace{-4pt}

There are two parts. First, the metric space(s) can be defined among (sub)sets, and therefore the distances can be defined using Hausdorff Distance; This can contribute to precision control, which can be used to adjust the value-information relationships reactively; and second, the existence test can be carried out via Cauchy Sequences (or its generalized methods).

\vspace{-8pt}

\subsection{Optimal Representations for One Monad.}

\vspace{-4pt}

This work also conjectures that: the optimal construction of the representation from one monad exists, and it can be derived via Cauchy Inequality (or its generalized methods). The optimal one is considered as the core values from one monad. Therefore, the transmission can be reduced significantly by transferring only the core values.

\vspace{-8pt}

\section{Implications on Information, Communication and Beyond}

\vspace{-4pt}

The above results deliver some implications on (information) communication, and these further impact the understanding of the intelligence (and our understanding on Monadology).

\vspace{-8pt}

\subsection{Functionality Agreements among Monads} 

\vspace{-4pt}

Rather than the focus on transmission, the conjecture on the existence of core values imply the need of functionality agreements: a descriptor of such an agreement can be used to restore all values, with the core values transmitted. Moreover, the descriptor can be surprisingly longer than the core values, and the transmission of the descriptor is intermittent (or, in other words, on demand).

\vspace{-8pt}

\subsection{Understanding on the Intelligence} 

\vspace{-4pt}

Given the co-existence of active and reactive actions presented in a human being, this work conjunctures that the Intelligence shall consist of \textit{at least} one active monad (who make choices) and one reactive monad (who accepts labels)\footnote{Such a conjecture is based on an interesting contradiction, made by the author on Gottfried Leibniz: why does the peak of rationalism at that time (1) accept the diversity of individuals; and (2) reject the diversity of possible worlds? Clearly, he made a choice to accept/reject something; and also received labels to accept/reject something.}
. Moreover, either form of a monad can reach to the optimal cognition, as long as its status is consistent: the analogy is the ZFC and NBG.

\vspace{-8pt}
    
\section{Conclusions}

\vspace{-8pt}

This work formalizes value, representation and information. With it, this work provides a new analytic framework, which allows quantitative methods to be used among (ordered) value, representation (if needed), and (disordered) information. The outstanding novelties are the suggestions on (1) value precision versus distance metrics; (2) existence test; and (3) optimal construction. Based on the above results, this work also revisits information and communication theory. 

\vspace{-8pt}

\end{singlespace}

\printbibliography

@article{BellTechn48/Shannon,
  title={{A Mathematical Theory of Communication}},
  author={Shannon, Claude Elwood},
  journal={{The Bell System Technical Journal}},
  volume={27},
  number={3},
  pages={379--423},
  year={1948},
  publisher={Nokia Bell Labs}
}

@article{ArXiv24/Peng,
  author = {Xiangjun Peng},
  title = {{Reversed Indexes {\(\approx\)} Values in Wavelet Trees}},
  year = {2023},
  journal = {ArXiv}
}

@book{leibniz/monadology,
  title={{The Monadology and Other Philosophical Writings}},
  author={Leibniz, Gottfried Wilhelm},
  year={1898}
}

@book{knuth/surreal-numbers,
  title={{Surreal Numbers}},
  author={Knuth, E Donald},
  year={1974},
  publisher={Addison-Wesley Professional}
}

@inproceedings{FOCS08/Succincter,
  author       = {Mihai P{u{a}}tra{c{s}}cu},
  title        = {{Succincter}},
  booktitle    = {{IEEE FOCS}},
  year         = {2008}
}

@book{leibniz-essais,
  title={{Essays of Theodicy on the Goodness of God, the Freedom of Man and the Origin of Evil}},
  author={Leibniz, Gottfried Wilhelm},
  year={1760}
}

\end{document}